\documentclass[aps,prb,showpacs,twocolumn]{revtex4-1}

\usepackage[english]{babel}
\usepackage[utf8]{inputenc}
\usepackage{amsmath}
\usepackage{amssymb}
\usepackage[caption=false]{subfig}
\usepackage{amssymb}
\usepackage{epsfig}
\usepackage{graphicx}
\usepackage{amsmath}
\usepackage{array,color}
\usepackage{natbib}

\usepackage[usenames,dvipsnames]{xcolor}
\definecolor{forestgreen}{rgb}{0.11,0.54,0.15}
\definecolor{purple}{rgb}{0.62,0.10,0.96}
\definecolor{dockerblue}{rgb}{0.11,0.56,0.98}
\definecolor{freeblue}{rgb}{0.25,0.41,0.88}

\usepackage[pdftex,plainpages=false,colorlinks=true,linkcolor=Red, citecolor=blue, urlcolor=blue]{hyperref}

\begin{document}

\title{Mott transition and magnetism on the anisotropic triangular lattice}
\author{S. Acheche$^1$, A. Reymbaut$^1$, M. Charlebois$^1$, D. S\'{e}n\'{e}chal$^1$ and A.-M. S. Tremblay$^{1,2}$}
\affiliation{
$^1$D\'{e}partement de physique, Institut Quantique \\ and Regroupement qu\'{e}b\'{e}cois sur les mat\'{e}riaux de pointe, Universit\'{e} de Sherbrooke, Sherbrooke, Qu\'{e}bec, Canada J1K 2R1 \\
$^2$Canadian Institute for Advanced Research, Toronto, Ontario, Canada, M5G 1Z8
}
\date{\today}
\begin{abstract}
Spin-liquid behavior was recently suggested experimentally in the moderately one-dimensional organic compound $\kappa$-H$_3$(Cat-EDT-TTF)$_2$. This compound can be modeled by the one-band Hubbard model on the anisotropic triangular lattice with $t’/t \simeq 1.5$, where $t'$ is the minority hopping. It thus becomes important to extend previous studies, that were performed in the range $0 \leq t'/t \leq 1.2$, to find out whether there is a regime where Mott insulating behavior can be found without long-range magnetic order. To this end, we study the above model in the range $1.2 \leq t'/t \leq 2$ using cluster dynamical mean-field theory (CDMFT). We argue that it is important to choose a symmetry-preserving cluster rather than a quasi one-dimensional cluster. We find that, upon increasing $t'/t$ beyond $t’/t \approx 1.3$, the Mott transition at zero-temperature is replaced by a first-order transition separating a metallic state from a collinear magnetic insulating state. Nevertheless, at the physically relevant value $t’/t \simeq 1.5$,  the transitions toward the magnetic and the Mott insulating phases are very close. The phase diagram obtained in this study can provide a working basis for moderately one-dimensional compounds on the anisotropic triangular lattice. 
\end{abstract}


\maketitle



\section{Introduction}
Organic superconductors of the BEDT family exhibit fascinating phenomena due to the interplay between strong electronic correlations and large magnetic frustration.\cite{PowellMcKenzieReview:2011,powell_strong_2006} For instance, their rich phase diagram contains $d$-wave superconducting, antiferromagnetic\cite{Lefebvre_2000} and possibly quantum spin liquid states at absolute zero.\cite{Shimizu_2003,Kurosaki_2005} At finite temperature, a Mott metal-insulator transition has clearly been identified.\cite{Lefebvre_2000} Recently, a moderately one-dimensional organic compound has been synthesized.\cite{Isono_2014} This 2D organic Mott insulator, $\kappa$-H$_3$(Cat-EDT-TTF)$_2$, does not exhibit magnetic order at very low temperature ($T = 50$ mK), which makes it a serious candidate for a quantum spin liquid.

Microscopically, the simplest model describing this organic compound is the two-dimensional single-band Hubbard model on an anisotropic triangular lattice.\cite{Isono_2014} This model is the same as that often used for BEDT organic compounds.\cite{TamuraBEDTdimer:1991, Kino_1996, McKenzie_1997, Jeschke_2012}

\begin{figure}[h!]
\begin{minipage}{1\linewidth}
\subfloat[]{\includegraphics[width=0.70\textwidth]{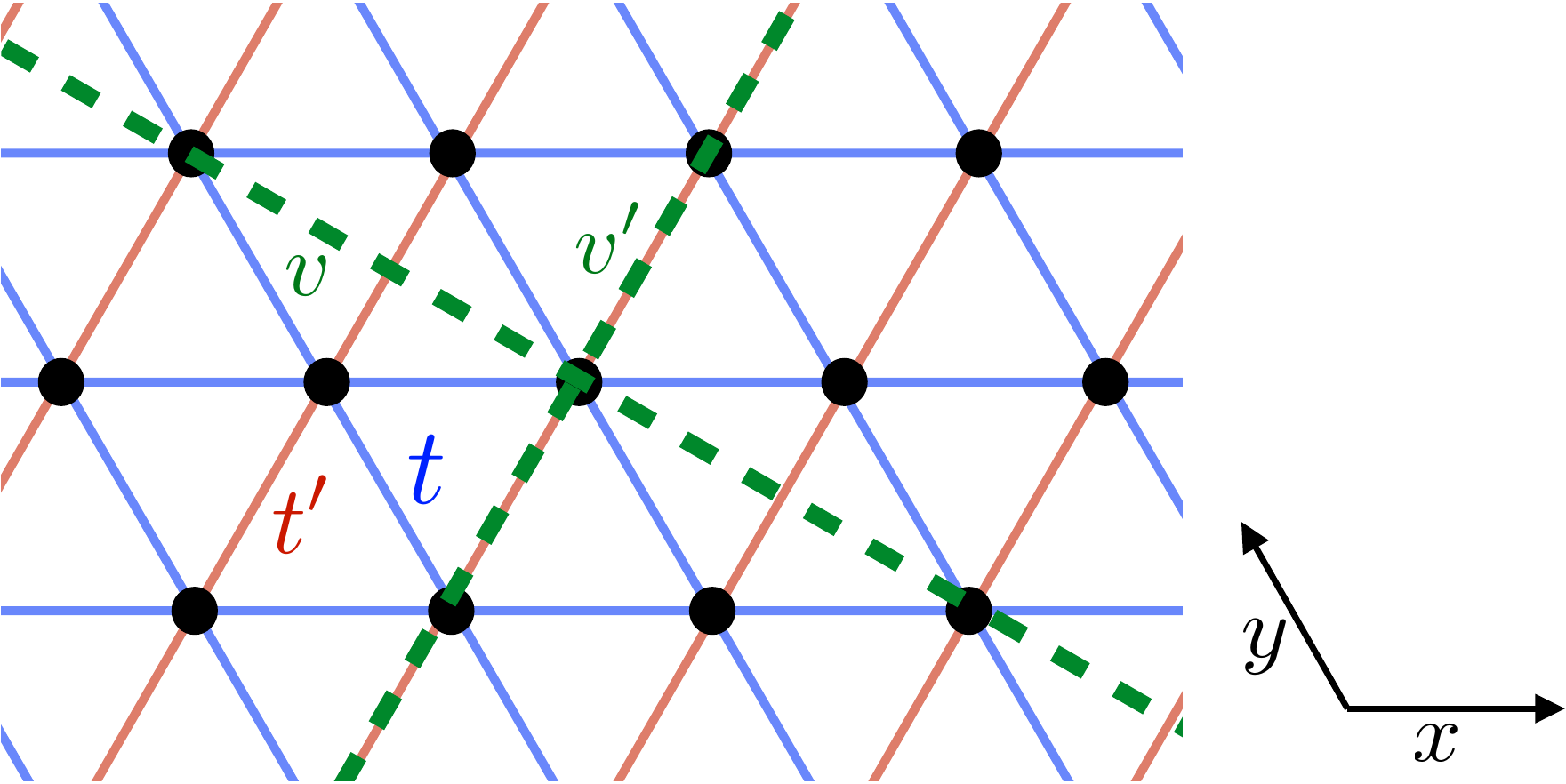} \label{Fig.sym_a}} 
\end{minipage}
\begin{minipage}{1\linewidth}
\subfloat[]{\includegraphics[width=0.90\textwidth]{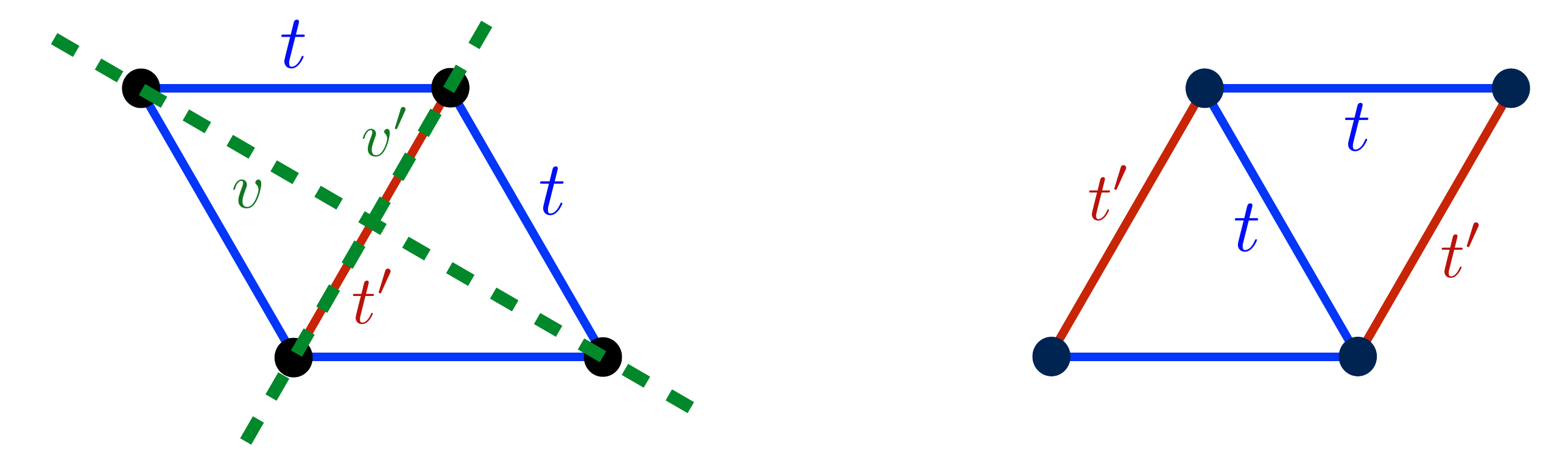} \label{Fig.sym_b}}
\end{minipage}
\centering
\caption{(Color online) (a) Illustration of the anisotropic triangular lattice with dashed green lines emphasizing the two mirror planes $v$ and $v'$. (b) While the first cluster geometry, called the symmetry-preserving (SP) cluster, displays the same symmetries $v$ and $v'$ as the infinite lattice, the second cluster geometry, called the quasi one-dimensional (Q1d) cluster, does not.}
\label{Fig.sym}
\end{figure}

In order to focus our attention on $\kappa$-H$_3$(Cat-EDT-TTF)$_2$, we study the phase diagram of this model in the moderately one-dimensional (M1d) regime characterized by a ratio between the hopping terms $t'$ and $t$, defined in Fig.\ref{Fig.sym_a}, larger than unity. Since we are interested in the Mott transition and the possibility of a spin liquid, we use cellular dynamical mean-field theory (CDMFT),\cite{Kotliar_2001, Lichtenstein_2000} a cluster extension of dynamical mean-field theory (DMFT) that can treat both the metallic and the insulating phases, the Mott transition between them as well as magnetic phases.\cite{Georges_1996}

Before we proceed, we briefly recall related studies, setting aside superconductivity that we do not consider here.  Previous work focused mostly on the frustrated regime $0 \leq t'/t \leq 1$ (square lattice to triangular lattice) since these anisotropy values, usually obtained from ab initio calculations,\cite{Kandpal_2009, Kazuma_2009,Nakamura_2012,Guterding_Valenti:2016, Tsumuraya_2015} corresponded to all known BEDT organic compounds. Theoretical investigations that were concerned with the Mott transition in the interaction-frustration ($U/t-t'/t$) phase diagram used methods that included path-integral renormalization group~\cite{MoritaBEDT:2002}, variational methods,~\cite{Watanabe:2006,Watanabe:2008} exact diagonalization,\cite{Clay:2008} Variational Cluster Approximation,~\cite{Sahebsara_2006} CDMFT \cite{Parcollet:2004,Kyung_2006,Kyung:2007, ohashi:2008} and dual-fermion~\cite{LeeDualFermions:2008} approaches. Although there are quantitative discrepancies between the different methods, metallic, insulating, non-magnetic, and antiferromagnetic phases were found, generally in good agreement with experiments. \cite{Kurosaki_2005, Shimizu_2003, Lefebvre_2000} A more detailed comparison between experiment and some of the above theoretical calculations appears in Ref.\onlinecite{Kandpal_2009}. CDMFT was one of the most successful approaches.\cite{Kyung_2006}

More recently, the M1d regime ($1\leq t'/t \leq 2$) has been investigated:\cite{Sahebsara_2008, Tocchio_2013} for $t'/t \approx 1$, the Hubbard model exhibits a spiral order and possibly a spin liquid phase. Different theoretical lattice approaches seem to agree with the presence of a spin liquid and a collinear magnetic state with an associated ordering vector $\textbf{Q} = (0, \pi)$ for $t'/t \geq 1.2$,\cite{Tocchio_2014, Yamada_2014} even though they do not completely agree on the precise form of the phase diagram, unlike for the case $t'/t<1$. In the strong-interaction limit, \emph{i.e} for $U \gg t,t'$, where the Hubbard model in the insulating phase reduces to the Heisenberg model with exchange interactions $J = 4t^2/U$ and $J' = 4t'^2/U$ to second order in perturbation theory\cite{MacDonald_1988} and, at higher order, to models with more complicated spin interactions, such as the two distinct ring exchange couplings $K = 80t^4/U^3$ and $K' = 80 t^2 t'^2/U^3$.\cite{Balents_2003}  The Heisenberg model corresponding to the M1d regime, \emph{i.e} $1\leq J'/J \leq 2$ has been extensively studied using different methods such as linear spin-wave,\cite{Merino_1999, Trumper_1999} coupled cluster method,\cite{Bishop_2009} variational Monte-Carlo\cite{Yunoki_2006,Heidarian_2009} or density matrix renormalization group.\cite{Weng_2006} These methods show that a spiral state is present for $J'/J \approx 1$ but they give different magnetic phases in the M1d regime, \emph{e.g} a spiral phase, a collinear magnetic phase or a spin liquid state. More sophisticated Hamiltonians including the ring exchange coupling $K$ in the anisotropic triangular lattice give a rich phase diagram where the presence of a Néel state, a spin liquid state or a spiral phase depends on the relative strength between $K/J$ and $J'/J$.\cite{Holt_2014}

This paper is organized as follows: the Hubbard model and the cellular dynamical mean-field theory (CDMFT) on a plaquette with an exact diagonalization (ED) impurity solver are detailed in Sec.\ref{Sec.model&method}. In Sec.\ref{Sec.results_Normal}, results for the normal state, showing a first order Mott metal-insulator transition are presented. Magnetic states are explored in Sec.\ref{Sec.results_Magnet}. Our results are summarized in the phase diagram of Fig.\ref{Fig.Entire_phase_diagram}. Finally, the choice of cluster is motivated in Sec.\ref{Sec_Cluster}, where we show, by comparing results for the magnetic phases with other methods, that a cluster sharing symmetries with the lattice is essential for a reliable CDMFT calculation in this regime. We conclude in Sec.\ref{Sec.conclusion}.

\section{Model and method}
\label{Sec.model&method}
We focus on the physics embodied in the Hubbard model on the anisotropic triangular lattice 
\begin{equation}
\hat{H} = -\sum_{i,j, \sigma} t_{ij}\, \hat{c}^\dagger_{i \sigma} \hat{c}_{j \sigma} + \, U \sum_i \hat{n}_{i \uparrow} \hat{n}_{i \downarrow} - \mu \sum_{i, \sigma} \hat{n}_{i \sigma}\, .
\end{equation}
All quantities are expressed in natural units ($\hbar \equiv 1$ and $k_B \equiv 1$). Here, $t_{ij}$ are the hopping amplitudes between sites $i$ and $j$ and can take two different values, $t$ and $t'$, illustrated in Fig.\ref{Fig.sym}. The Fourier transform of the hopping amplitudes $t_{ij}$ determines the anisotropic bare dispersion $\epsilon_{\textbf{k}} = -2t\, [\cos(k_x) + \cos(k_y)] -2t'\cos(k_x + k_y)$. The on-site Coulomb repulsion is $U$ and $\mu$ is the chemical potential set so that the system is half-filled. For that filling, the signs of $t$ and $t'$ do not modify the phase diagram. Layered organic compounds are usually half-filled, but doped compounds~\cite{Lyubovskaya-1:1987,Lyubovskaya-2:1987} have been investigated experimentally~\cite{Oike_2015,Oike:2016} and theoretically.\cite{Watanabe_SC_2014,hebert_superconducting_2015}

We focus on the M1d regime $1.2 \leq t'/t \leq 2$ using CDMFT,\cite{Kotliar_2001, Lichtenstein_2000} a cluster extension of dynamical mean-field theory (DMFT).\cite{Georges_1996} CDMFT approximates the infinite lattice as a finite size cluster self-consistently coupled to a bath of non-interacting electrons, thus taking into account dynamical correlations as well as spatial correlations up to the size of the cluster.\cite{Maier_2005, Senechal_2010} CDMFT, like DMFT, maps the system into an Anderson impurity problem, which is then solved self-consistently. In this paper, the quantum impurity problem is solved using the exact diagonalization (ED) method \cite{Caffarel_1994} at zero temperature. This method is restricted to a small number $N_b$ of bath sites. While the Hamiltonian of the quantum impurity problem is coded exactly, the ground state and the Green functions of interest are found in a quasi-exact way with the Lanczos algorithm.\cite{Senechal_2010} Exact diagonalization is robust in the presence of frustration, unlike quantum Monte-Carlo methods which suffer from the infamous sign problem.\cite{Troyer_2005} Moreover, it directly computes dynamical quantities as a function of real frequencies. To summarize, the assumption inherent to cluster approaches is that the essential physics of the system originates from short-ranged correlations, which should be the case in strongly-correlated magnetically-frustrated organic compounds. 

We solve the following cluster-bath Hamiltonian:
\begin{eqnarray}\label{Eq.model}
\nonumber
\hat{H} &=& \sum_{i, j, \sigma} t_{ij} \hat{c}^\dagger_{i \sigma} \hat{c}_{j \sigma} + U \sum_{i} \hat{n}_{i \uparrow} \hat{n}_{i \downarrow} -\mu \sum_{i,\sigma} \hat{n}_{i \sigma}  \\
\nonumber
&+& \sum_{m, \sigma} \epsilon_{m \sigma} \hat{b}^{\dagger}_{m \sigma} \hat{b}_{m \sigma}\\
&+& \sum_{m, i, \sigma} \theta_{m i \sigma} (\hat{b}^{\dagger}_{m \sigma} \hat{c}_{i \sigma} + h.c)\, ,
\end{eqnarray}
where the indices $i,j = 1, \ldots, N_c$ label the sites within the cluster whereas $m = 1, \ldots, N_b$ label the bath sites. The second quantized operators $\hat{c}_{i \sigma}$ and $\hat{b}_{m \sigma}$ annihilate electrons on the cluster and in the bath, respectively. $t_{ij}$ are the hopping matrix elements within the cluster, $\epsilon_{m \sigma}$ are the bath energies, and $\theta_{m l \sigma}$ are the bath-cluster hybridization matrix elements. Besides, in order to allow antiferromagnetism to appear, $\epsilon_{m \sigma}$ and $\theta_{m l \sigma}$ explicitly carry a spin variable $\sigma$. A complication of the ED method is that the CDMFT self-consistency condition cannot be satisfied exactly because of the finite number of bath sites. This condition is rather satisfied approximately by minimizing a distance function. For further information on the matter, see  Refs.~\onlinecite{Caffarel_1994}, \onlinecite{Senechal_2010_b} and \onlinecite{Charlebois_2015}. We use the same distance function parameters as in the last two references, namely a frequency cut-off at $\omega_n/t = 2$ and a "fictitious" inverse temperature $\beta/t = 50$. To check that there are no artifacts associated with the finite bath, we checked our results for the Mott transition using CDMFT with a continuous-time quantum Monte-Carlo (CTQMC) impurity solver \cite{Haule_2007, Gull_2011} at finite but low temperature ($\beta/t = 20$). For both clusters that we describe below, we found agreement with our CDMFT plus ED solver for the values of $t'/t$ that we tested (0.4 and 1.5).\footnote{We thank P. S\'emon for providing us the impurity solver for the CTQMC calculation.}

\begin{figure}[h!]
\subfloat[]{\includegraphics[width=0.22\textwidth]{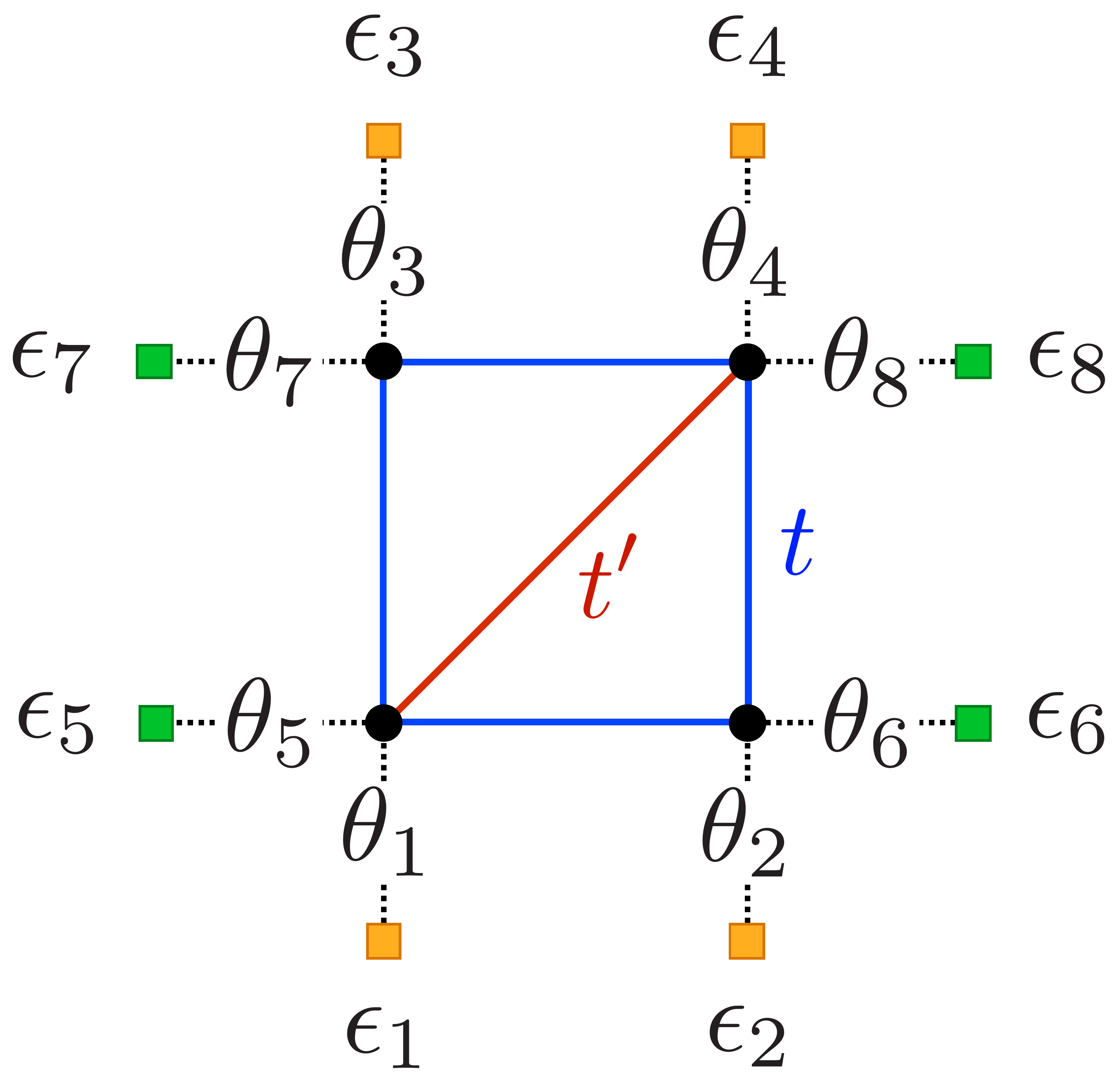} \label{Fig_SP}}
\hspace*{5pt}
\subfloat[]{\includegraphics[width=0.22\textwidth]{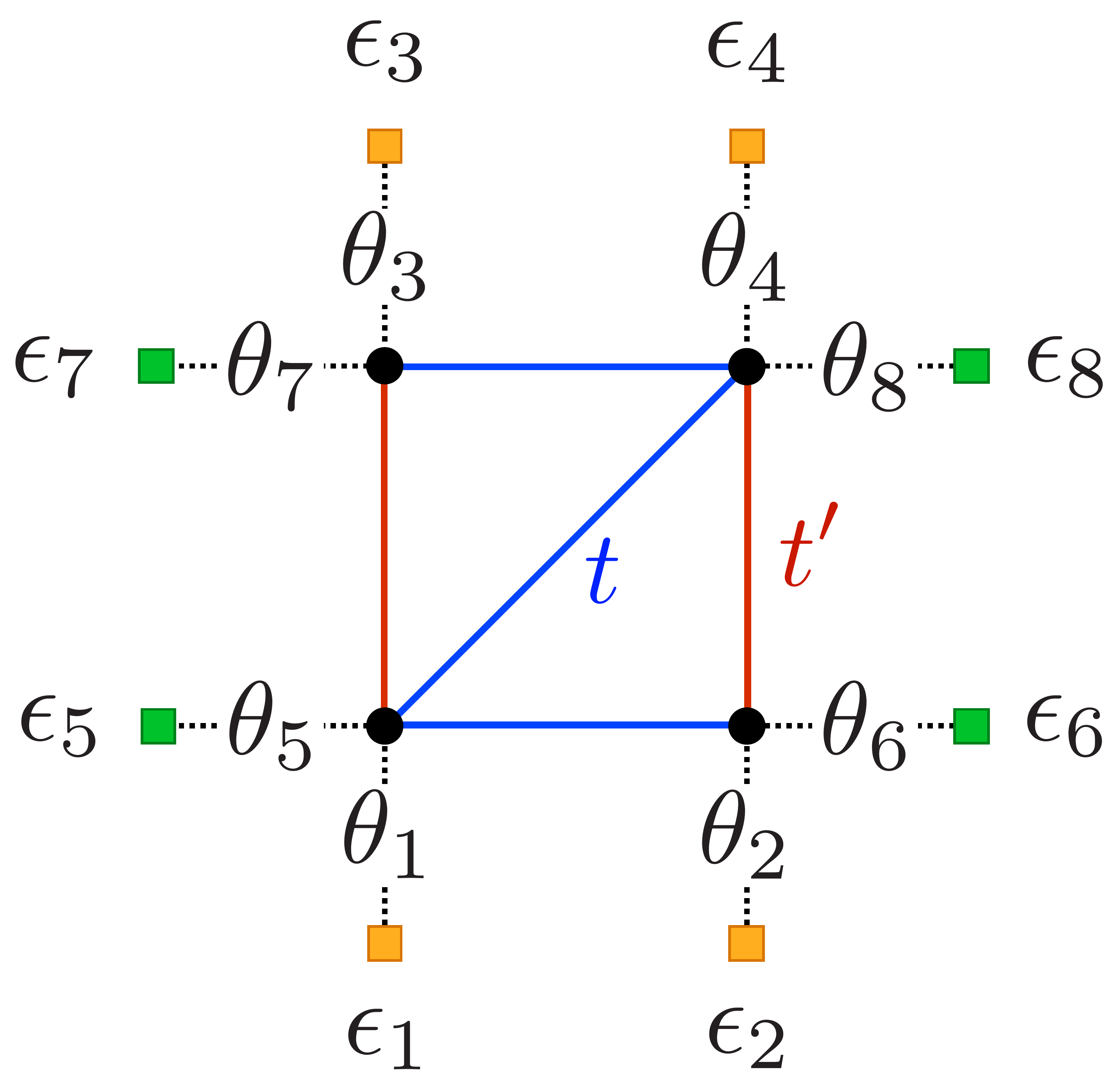} \label{Fig_Q1d}}
\centering
\caption{(Color online) (a) Symmetry preserving (SP) cluster. (b) Quasi-one-dimensional (Q1d) cluster. The four cluster sites are black circles and the bath sites are green and orange squares. For simplicity, spin indices ${\sigma}$ for the bath energies $\epsilon_{m,\sigma}$ and cluster and spin indices ${i,\sigma}$ for the bath-cluster hybridization matrix elements $\theta_{m,i,\sigma}$ are not explicitly shown. Depending on the symmetry of the phase being explored, some of the variational parameters are taken equal. Note that the bath sites do not have a position in real space and that the Q1d cluster does not share the symmetries of the lattice, contrary to the SP cluster.}
  \label{cluster_cartoon}
\end{figure}

As illustrated in Fig.\ref{cluster_cartoon}, we use clusters of $N_c=4$ sites coupled to $N_b = 8$ bath sites. Although the calculation is for 2$\times$2 clusters, we expect to capture the main physics of the lattice since studies using CDMFT have confirmed that results on a 2$\times$2 cluster are quantitatively similar to those on larger clusters, at least at high temperature.\cite{Sakai_2012} All physical results presented in the next sections are extracted from the symmetry-preserving (SP) cluster of Fig.\ref{Fig.sym_b}, whose parametrization within the model Eq.\eqref{Eq.model} is detailed in Fig.\ref{Fig_SP}. For large values of $t'/t$, one might argue that the quasi one-dimensional character of the lattice must be present in the cluster. In order to shed light on this question, the quasi-one dimensional (Q1d) cluster of Fig.\ref{Fig.sym_b}, whose parametrization within the model Eq.\eqref{Eq.model} is detailed in Fig.\ref{Fig_Q1d}, has also been investigated. Our results, presented in the following sections, will show that the physics extracted from this second cluster geometry does not compare well with other methods, leading us to conclude that the SP cluster is a better representation of the infinite lattice in the M1d regime.

\section{Results}

\subsection{Mott transition} 
\label{Sec.results_Normal}

\begin{figure}[h!]
\subfloat[]{\includegraphics[width=0.22\textwidth]{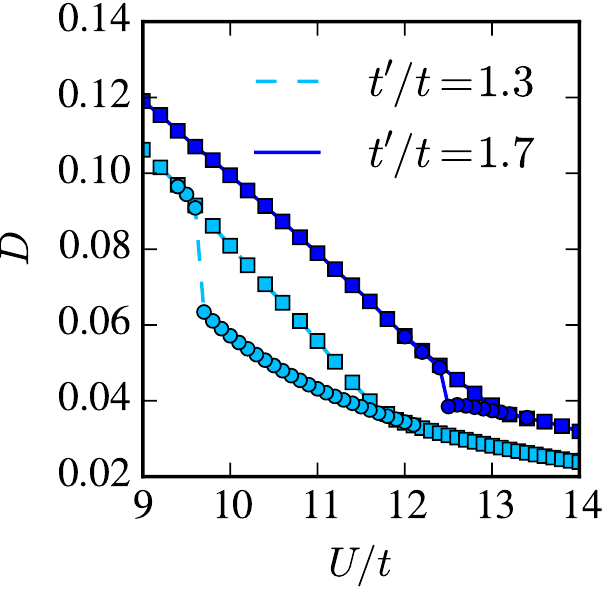} \label{D}}
\hspace*{9pt}
\subfloat[]{\includegraphics[width=0.22\textwidth]{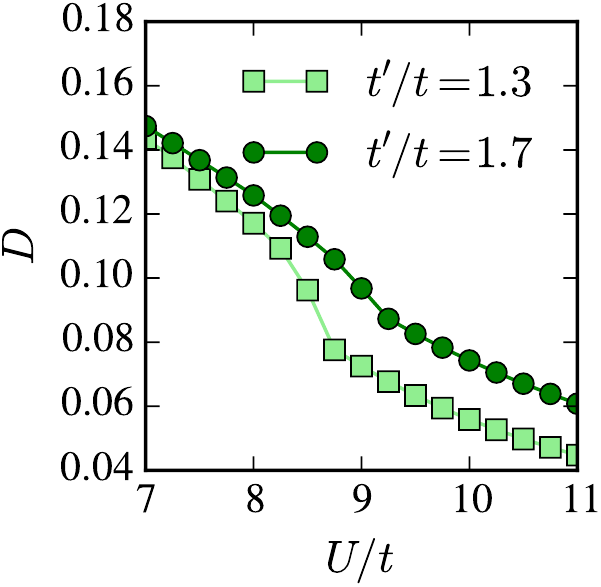} \label{D_NG}}\\

\vspace*{5pt}

\begin{minipage}{1\linewidth}
\subfloat[]{\includegraphics[width=0.85\textwidth]{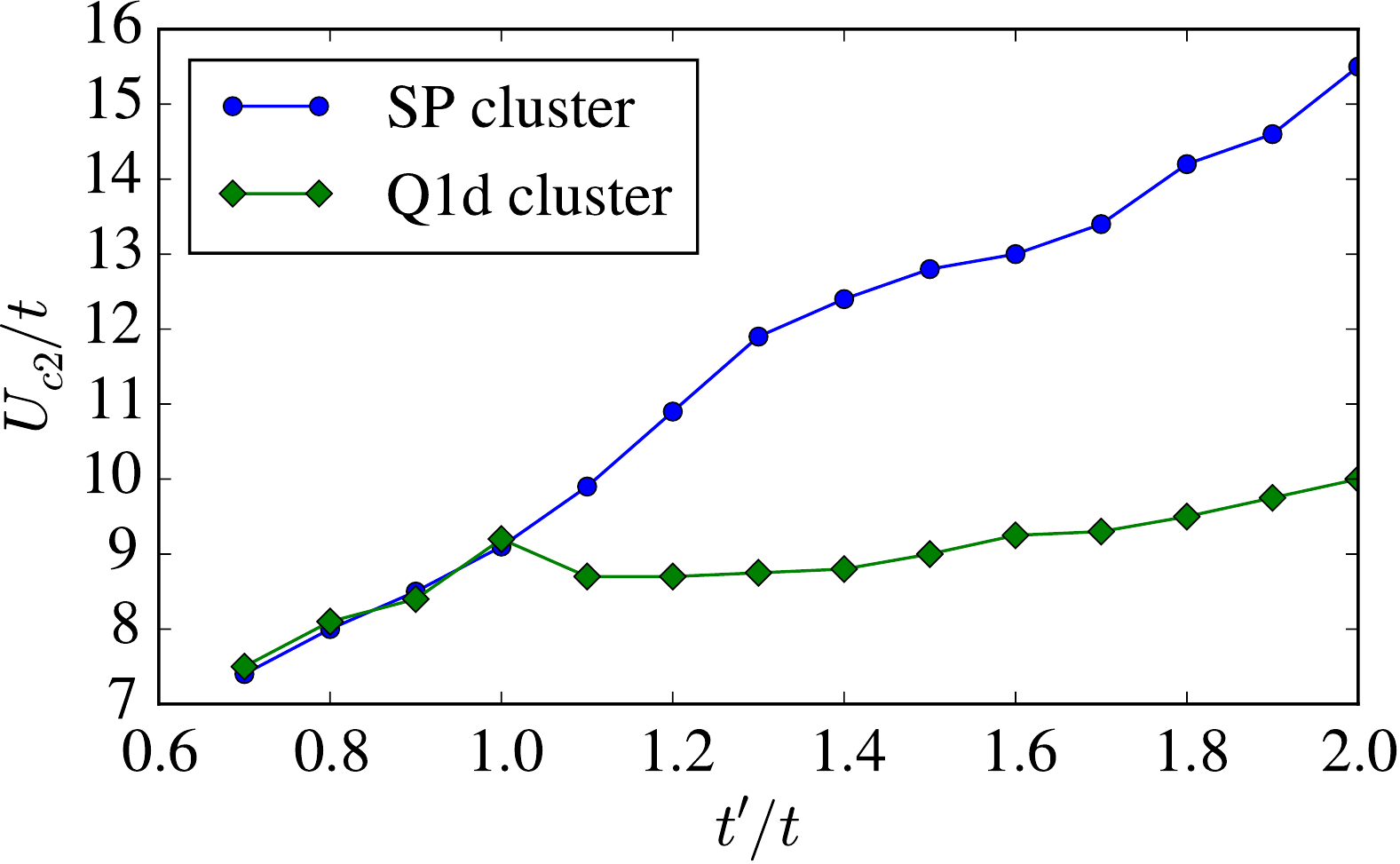}\label{Fig.comparaison}}
\end{minipage}
\centering
\caption{(Color online) (a) Double occupancy $D = \langle \hat{n}_\uparrow \hat{n}_\downarrow \rangle$ for the SP cluster as a function of $U/t$ for $t'/t = 1.3$ (sky blue dashed line) and $t'/t = 1.7$ (dark blue solid line). While the jump at low interaction defines the lower critical ratio $U_{c1}/t$, the change of slope at higher interaction defines the upper critical ratio $U_{c2}/t$. (b) Double occupancy for the Q1d cluster as a function of $U/t$ for $t'/t = 1.3$ (light green solid line) and $t'/t = 1.7$ (dark green solid line). Here, the change of slope defines a critical ratio $U_{c2}/t$ analog to the upper critical ratio of the SP cluster. It is investigated further through the low-frequency behavior of the local density of states, Fig.\ref{Fig.DOS}. (c) Mott critical ratio $U_{c2}/t$ as a function of $t'/t$ for the SP and Q1d clusters, in dark blue and dark green, respectively. For $t'/t \leq 1$, the results are quantitatively equivalent but differences clearly appear just above $t'/t=1$.}
  \label{fig:magn}
\end{figure}

\begin{figure}[h!]
\begin{minipage}{1 \linewidth}
\includegraphics[width = 1\textwidth]{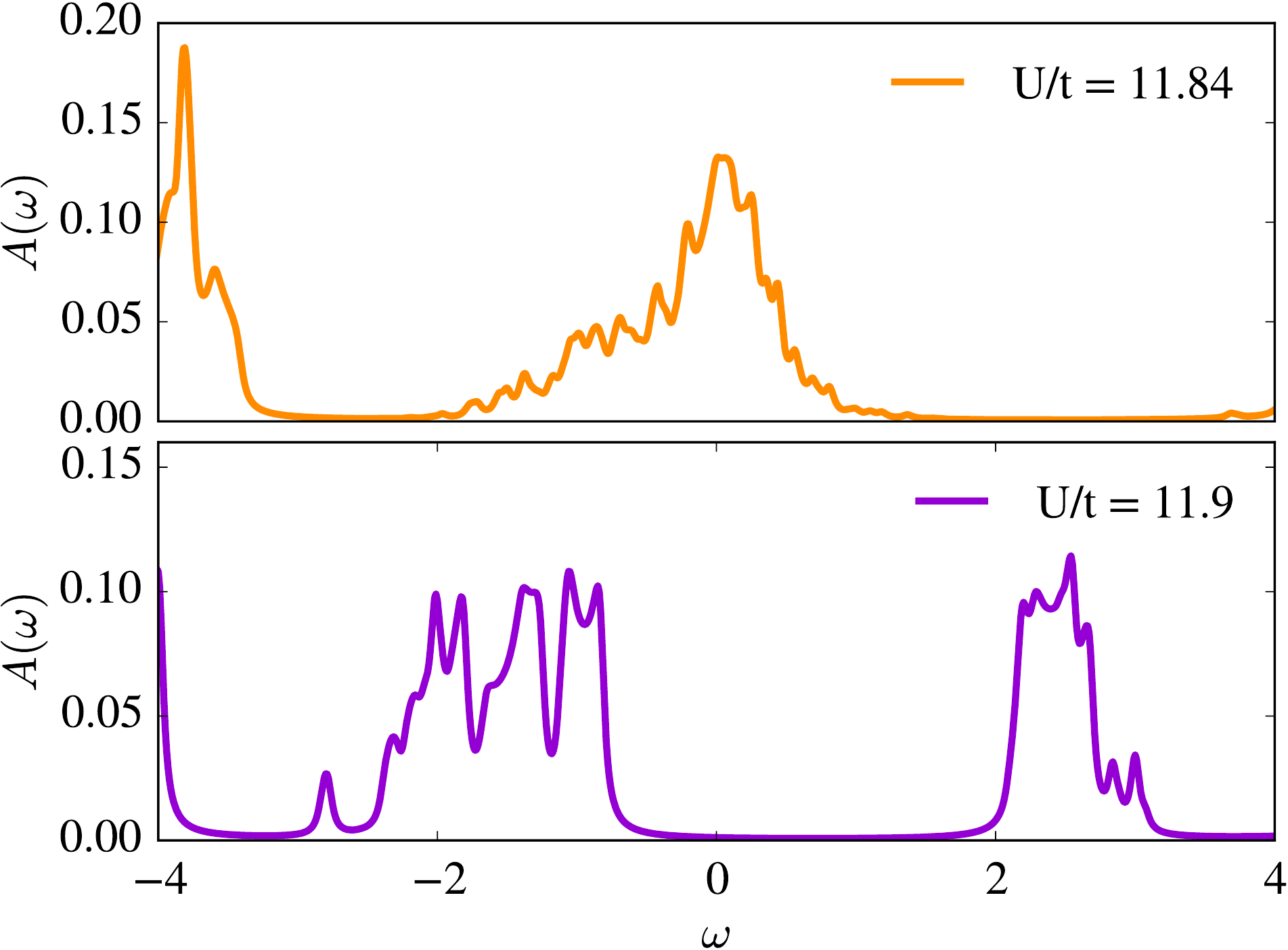}
\end{minipage}
\caption{(color online) Local density of states $A(\omega)$ for $t'/t = 1.3$ in the normal state at $U/t = 11.84$ and $U/t = 11.9$ on the SP cluster. A Lorentzian broadening $\eta = 0.035$ was used. A metal-insulator transition occurs between these two values of $U/t $.}
\label{Fig.DOS}
\end{figure}

First, let us focus on the Mott transition in the normal state. Magnetic states are forbidden if one suppresses the spin dependence of the bath parameters. The phase diagram was obtained by changing $t'/t $ in steps of $0.1$ and varying $U/t$ in much smaller steps. Therefore, the values of $t'/t $ quoted for phase boundaries in the following have an uncertainty of order $\pm 0.05$. We find that as long as $t'/t \leq 1.2$, the double occupancy displays hysteresis bounded by jumps at $U_{c1}$ and $U_{c2}$ as $U$ decreases or increases, respectively ($U_{c1} \leq U_{c2}$). These jumps are the signature of the usual Mott transition. For the SP cluster at $t'/t \geq 1.3$, the hysteresis region is still present but is bounded by a jump only when $U$ decreases. As $U$ increases, a mere change of slope occurs, as shown in Fig.\ref{D}. Even without a jump in the double-occupancy, the Mott transition can be observed by studying the low-frequency behavior of the local density of states $A (\omega)$, as presented in Fig.\ref{Fig.DOS} for $t'/t = 1.3$. For $U/t = 11.84$, slightly smaller than the upper critical ratio $U_{c2}/t$ at $t'/t = 1.3$, the local density of states exhibits a low-frequency metallic behavior. When $U/t$ is increased only by a tiny fraction to $U/t = 11.9$, the opening of an insulating gap in the local density of states signals the Mott transition. The Mott transition of the Q1d cluster, presented in Fig.\ref{D_NG}, does not even feature a jump or some hysteresis in the double occupancy. The local density of states, however, indicates a Mott transition (not shown). 

The critical ratios $U_{c2}/t$ presented in Fig.\ref{Fig.comparaison} as a function of $t'/t$ illustrate one of the main differences between the two cluster geometries considered in this paper. Indeed, for $t'/t > 1$, $U_{c2}/t$ first decreases before increasing for the Q1d cluster whereas it only increases monotonically for the SP cluster. Surprisingly, this discrepancy between the two cluster geometries does not hold for $t'/t \leq 1$ since both geometries yield the same value of $U_{c2}/t$ even if one could assume that the Q1d cluster should be more appropriate only in the M1d regime. An acceptable explanation for this phenomenon has not been found yet, but the following section will give arguments that lead us to believe that the results for the SP cluster capture the correct physics. 

\subsection{Magnetic states}
\label{Sec.results_Magnet}

Within CDMFT, one can only look for commensurate magnetic orders on the cluster. Therefore, this restriction does not allow us to explore all possible magnetic phases nor to distinguish between a spin liquid and an incommensurate magnetic order (in the sense of a magnetic order whose unit cell does not perfectly fit or repeat within the cluster). Hence, we can only rule out a spin liquid by demonstrating that a magnetic phase exists, but we cannot prove that a spin liquid state will occur since we cannot explore all possible magnetic states. In other words, not finding one of the allowed magnetic states of our cluster in a Mott insulating phase is a necessary, but not sufficient, condition for a spin liquid. A spin-liquid state could occur only in a non-magnetic insulating state (NMI state).  

\begin{figure}[h!] 
\center
\begin{minipage}{1 \linewidth}
\subfloat[]{\includegraphics[width=0.95\textwidth]{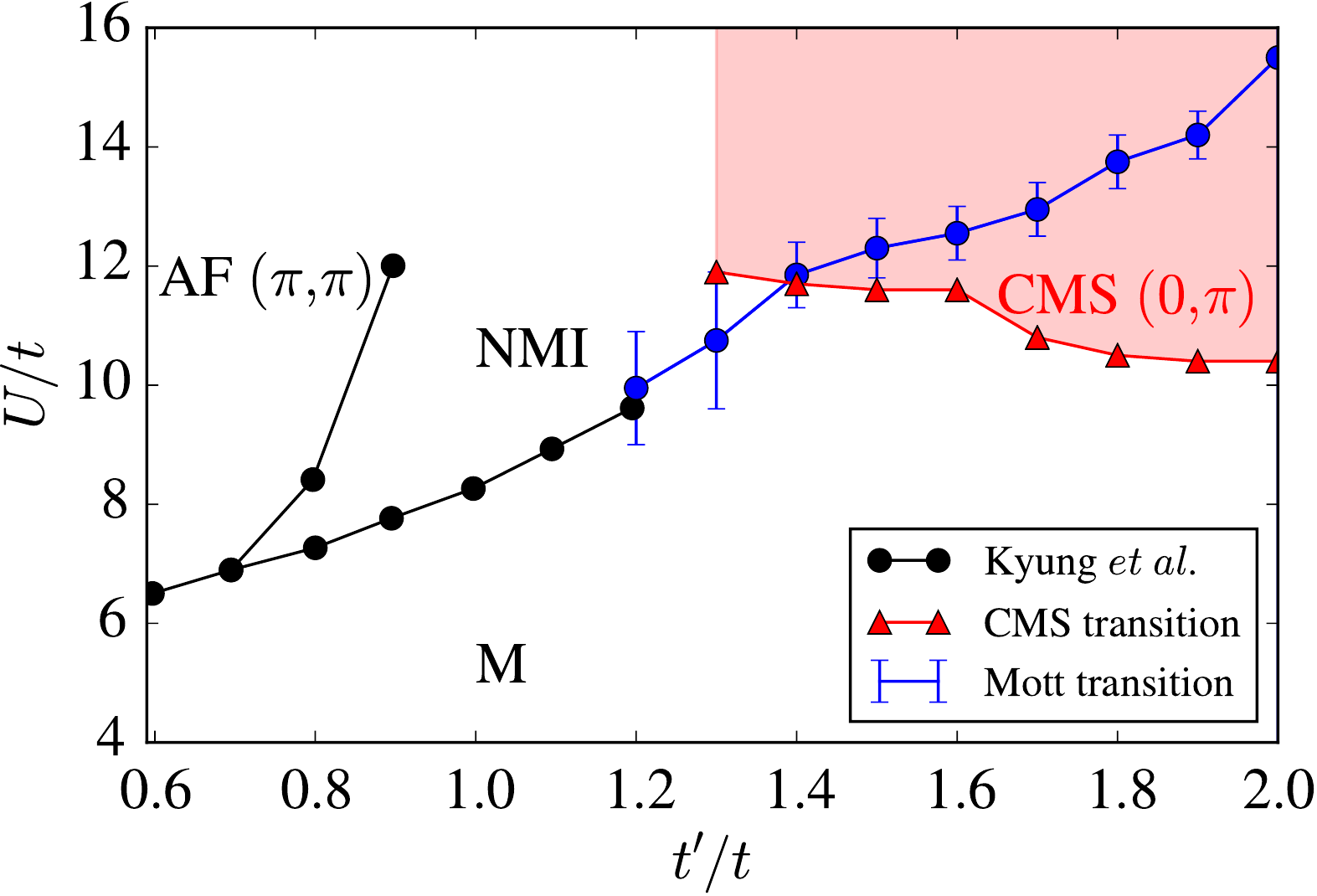}\label{Fig.Phase_diagram}}
\end{minipage}
\centering
\begin{minipage}{1 \linewidth}
\subfloat[]{\label{Fig.Tocchio}\includegraphics[width=0.96\textwidth]{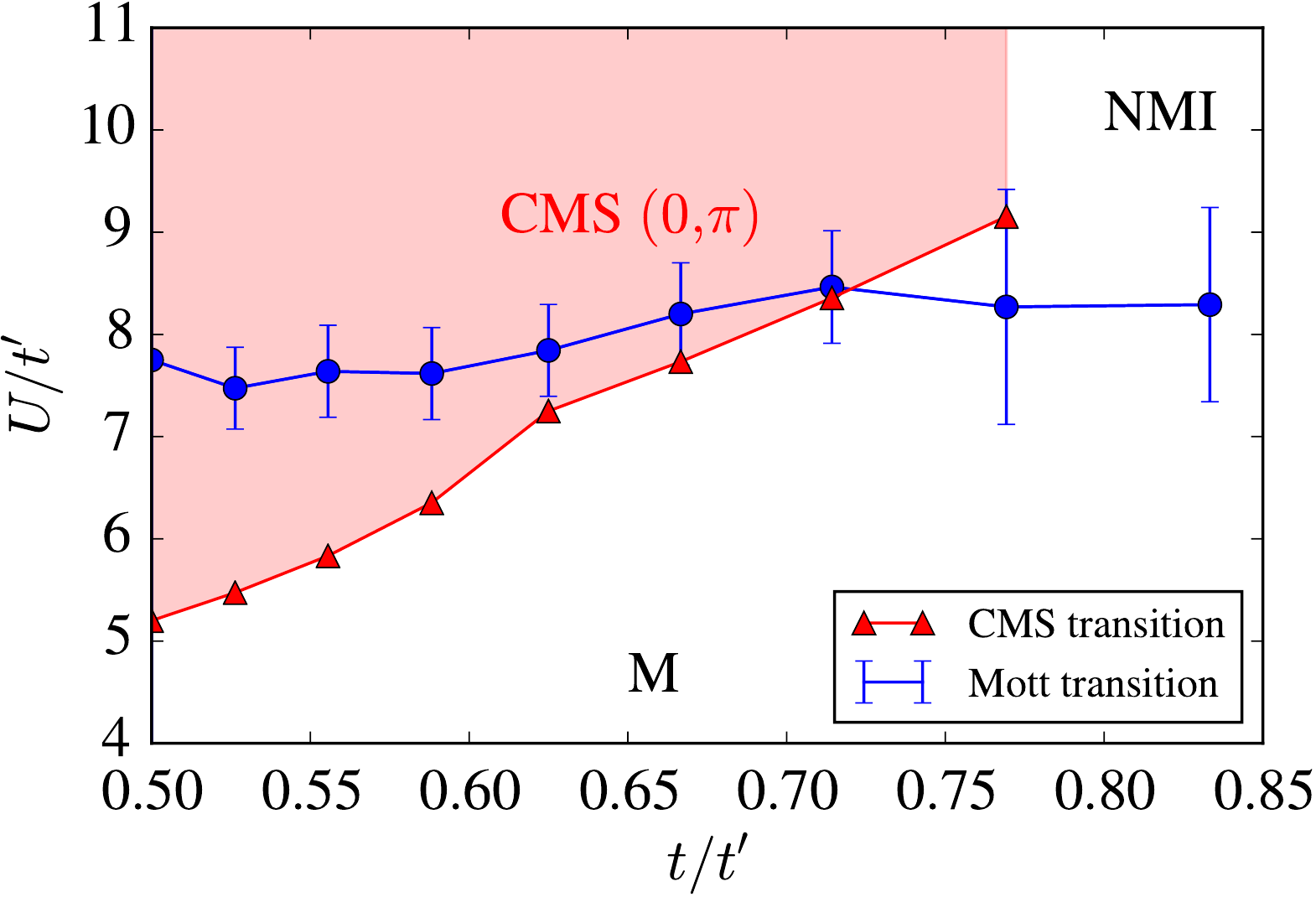}\label{Fig.Phase_diagram_alt}}
\end{minipage}

\caption{(Color online) (a) Phase diagram for the Hubbard model obtained with CDMFT on the SP cluster. Black curve: Data from a previous study of the low-frustration phase diagram carried out by Kyung \textit{et al.} (Ref.\onlinecite{Kyung_2006}). Blue curve: Mott transition in the non-magnetic normal state. The bars indicate the boundaries of the coexistence region. Red triangles: Metal to collinear magnetic state transition. Here, only $U^{\mathrm{CMS}}_{c2}/t$ is displayed. The lower critical interaction $U^{\mathrm{CMS}}_{c1}/t$ cannot always be found due to some numerical instabilities. Red area: The collinear magnetic phase with wave-vector $\textbf{Q} = (0, \pi)$. AF, NMI, CMS, and M denote the N\'eel state, the non-magnetic insulator, the collinear magnetic state and the metallic state, respectively. (b) Same phase diagram with $t'$ as energy unit, namely $U/t'$ vs $t/t'$ for $1.2<t'/t<2$. This phase diagram can be more easily compared with the results of Ref.\onlinecite{Tocchio_2014}.}
\label{Fig.Entire_phase_diagram}
\end{figure}

For reasons that will become clear below, we present our final phase diagram, including magnetic order, only for the SP cluster. One can check from Fig.\ref{Fig.Phase_diagram} that for $t'/t \leq 1.2$, we find the same results as in Ref.\onlinecite{Kyung_2006}, namely a transition from a metal to a N\'eel state for $t'/t \leq 0.7$, followed by a NMI state that starts right above the Mott transition for $t'/t \approx 0.7$ and then undergoes a N\'eel transition at larger $U/t$ if $0.7 \lesssim t'/t \lesssim 0.9$. We did not investigate $U/t > 12$. Previous studies indicate that a spiral order or a spin liquid could be present in this area of the phase diagram,\cite{Sahebsara_2008, Watanabe_2008, Laubach_2015} corresponding to the NMI state of Fig.\ref{Fig.Phase_diagram}. 

For $t'/t\geq 1.3$, a first-order transition between a metal and a collinear magnetic insulating state (CMS) takes place for $(0, \pi)$ or $(\pi,0)$ upon increasing $U/t$,  before the Mott transition, as shown in Fig.\ref{Fig.Phase_diagram}, and survives at larger values of $U/t$. The presence of this phase is not surprising since different studies predict the appearance of this magnetic phase in the M1d regime using lattice models \cite{Tocchio_2014, Yamada_2014} or spin models.\cite{Bishop_2009, Starykh_2010, Weng_2006, Holt_2014} At this magnetic transition, we observe a jump in the double-occupancy and a gap opening in the spectral function. Some hysteresis can be seen in the double-occupancy, but while the upper critical interaction $U_{c2}^{\mathrm{CMS}}$ can always be detected for any value of $t'/t$, the lower critical interaction $U_{c1}^{\mathrm{CMS}}$ cannot always be found due to some numerical instability (hence the absence of bars for the red triangles of Fig.\ref{Fig.Phase_diagram}). Fig.\ref{Fig.Phase_diagram_alt} shows the same phase diagram as Fig.\ref{Fig.Phase_diagram} using $t'$ instead of $t$ as the unit of energy to allow an easier comparison with the results of Ref.\onlinecite{Tocchio_2014}.

It is interesting to compare the phase diagram of Fig.\ref{Fig.Phase_diagram} with the one presented in Ref.\onlinecite{Nevidomskyy_2008} for the half-filled square lattice with nearest-neighbor hopping (more suited to the study of cuprates). This model is different and the method is the variational cluster approximation (VCA),\cite{Potthoff:2003} but it presents the same collinear magnetic phase with ordering wave vector $(0, \pi)$ or $(\pi,0)$, for $t'/t$ larger than $t'/t\approx 0.9$.

\section{Choice of CDMFT cluster}\label{Sec_Cluster}

We saw that the results obtained with the Q1d cluster differ from those for the SP cluster.  We have checked that the assumption that the Mott transition occurs when $U$ is of the order of the bandwidth is not sufficient to choose the appropriate cluster, although in this context the non-monotonic dependence on $t'/t$ of the Mott metal-insulator transition is suggestive of the inadequacy of the Q1d cluster. However, based on the symmetries satisfied by the SP cluster, as illustrated in Fig.\ref{Fig.sym}, it should capture the correct physics. This is confirmed by the fact that the predictions for the magnetic state obtained with the SP cluster agree with the results of other methods that are available for comparisons. With the Q1d cluster there is no commensurable magnetic state at all in the M1d regime while other methods find stable magnetic long-range order. 

First, we searched for collinear magnetic state $(0, \pi)$ using the restricted Hartree-Fock approximation on a 18$\times$18 cluster. This method allows one, in principle, to map the phase diagram for a large but finite number of magnetic states and to study larger clusters than in CDMFT. Here, we used Hartree-Fock just to confirm our magnetic phase diagram in the M1d regime for the SP cluster. We thus allowed only two magnetic states: the Néel order and the collinear magnetic state. Even though CDMFT and Hartree-Fock methods cannot be compared quantitatively, a qualitative agreement is found: for $0 < t'/t \lessapprox 1$, a first-order metal to antiferromagnetic ($\pi$, $\pi$) insulator transition takes place at a finite interaction $U_c/t$, whereas for $t'/t \geq 1.2$, a first order metal to collinear magnetic insulator transition is found.  Although the same magnetic states and the same order of transition are found in the same range of $t'/t$ as in the CDMFT plus ED solver method, the critical Hartree-Fock interaction $U_c/t$ has a lower value, around $U_c/t\approx 6$, mainly due to the fact that the Hartree-Fock method is purely a mean-field theory that neglects the fluctuations that renormalize down the value of the interaction. Kanamori-Br\"uckner screening is an example of renormalization mechanism that is neglected in Hartree-Fock. 

Finally, the critical interaction for collinear magnetism found with the SP cluster exhibits qualitatively the same dependency on frustration as in the phase diagram of Ref.\onlinecite{Tocchio_2014} obtained by variational methods. The phase diagram in Fig.\ref{Fig.Tocchio} presents our results with the same axis as in Ref.\onlinecite{Tocchio_2014} to ease the comparison. 

Even though it is not a rigorous proof, the fact that the appearance of a collinear magnetic state for $1.3 < t'/t < 2$ is supported by three different numerical methods gives solid arguments in favor of its presence in this region of the phase diagram.  The lack of collinear magnetism with the Q1d cluster is an additional argument, beyond symmetry, to reject that cluster.

\section{Conclusion}
\label{Sec.conclusion}

In the moderately one-dimensional regime of the Hubbard model on the anisotropic triangular lattice, a symmetry-preserving cluster should be preferred to a quasi one-dimensional cluster geometry for calculations with cluster dynamical mean-field theory. The symmetry-preserving cluster gives magnetic phases in agreement with other methods.   

With the symmetry-preserving cluster, we obtained the phase diagram using CDMFT with an exact diagonalization solver in the moderately one-dimensional regime. There is a line of first-order Mott transition where the critical $U/t$ monotonically increases with $t'/t$. We also found a first-order metal-to-collinear magnetic state transition that occurs for $t'/t \geq 1.3$ and does not allow any spin liquid state to appear in this regime. For $0.7 \leq t'/t \leq 1.2$, no sign of magnetic states covering the metal-insulator transition has been found. A spin liquid or a magnetic order which is not commensurate with our cluster, such as a spiral order, might however appear in this region.  

Our results at $t'/t \simeq 1.5$ are particularly relevant for experiment since they are supposed to describe the organic compound $\kappa$-H$_3$(Cat-EDT-TTF)$_2$ that seems to be a good candidate for a spin liquid state.\cite{Isono_2014} At $t'/t \simeq 1.5$, the collinear magnetic state appears at a critical interaction strength that is slightly lower than the Mott critical interaction in the normal state. This would suggest that there cannot be a spin liquid state, \textit{i.e.} a non-magnetic insulating state, since the magnetic phase covers the Mott metal-insulator transition. However, the uncertainty on the value of $t'/t$ and the fact that the magnetic and Mott transitions are so close to each other does not allow us to strictly exclude a spin liquid state for $t'/t \simeq 1.5$. Had $t'/t$ been much larger, this possibility would have been excluded. One should also recall that other physical effects may need to be taken into account to model the real material, such as multi-band effects, near-neighbor repulsion etc. 


\section{Acknowledgments} We are grateful to R. Valenti, S. Verret and especially M. Gringas, for numerous suggestions and discussions. We also thank P. S\'emon for the CTQMC impurity solver. This work has been supported by the Natural Sciences and Engineering Research Council of Canada (NSERC) under grant RGPIN-2014-04584, and by the Research Chair in the Theory of Quantum Materials at Universit\'e de Sherbrooke (A.-M.S. T.). Simulations were performed on computers provided by Canadian Foundation for Innovation, the Minist\`ere de l’\'Education des Loisirs et du Sport (Qu\'ebec), Calcul Qu\'ebec, and Compute Canada.

\end{document}